\documentclass[twocolumn,prl,floatfix,superscriptaddress]{revtex4}
\usepackage{graphics}
\usepackage{epsfig}

\begin{document}

\title
{Excitonic Effects and Optical Spectra of Single-Walled Carbon Nanotubes} 
\author{Catalin D. Spataru} 
\author{Sohrab Ismail-Beigi}
\affiliation{Department of Physics, University of California at Berkeley,
Berkeley, California 94720}
\affiliation{Materials Sciences Division, Lawrence Berkeley National
Laboratory, Berkeley, California 94720}
\author{Lorin X. Benedict}
\affiliation{H Division, Physics and Advanced Technologies
Directorate, Lawrence Livermore National Laboratory, University of California, 
Livermore, CA 94550}
\author{Steven G. Louie}
\affiliation{Department of Physics, University of California at Berkeley,
Berkeley, California 94720}
\affiliation{Materials Sciences Division, Lawrence Berkeley National
Laboratory, Berkeley, California 94720}

\date{\today}

\begin{abstract}
Many-electron effects often dramatically modify the properties of
reduced dimensional systems. We report calculations, based on
an {\it ab initio} many-electron Green's function approach, of
electron-hole interaction effects on the optical spectra of
small-diameter single-walled carbon nanotubes. Excitonic effects
qualitatively alter the optical spectra of both semiconducting and
metallic tubes. Excitons are bound by $\sim 1$ eV in the
semiconducting (8,0) tube
and by $\sim 100$ meV in the metallic (3,3) tube. These large
many-electron effects explain the discrepancies between previous theories
and experiments.
\end{abstract}
\maketitle

Synthesis and observation of single-walled carbon nanotubes (SWCNT)
have advanced greatly in recent years, making possible the
experimental study of the optical properties of individual SWCNTs
\cite{ZMLi,OConnell}. If well understood, the optical response of
SWCNTs may be used to characterize these nanotubes, to monitor and
guide their separation by type \cite{Bachilo}, and can be employed in
device applications \cite{Misewich}. However, measured optical
transition frequencies deviate substantially from theoretical
predictions based on one-particle interband theories. This deviation
is not unexpected since many-body interactions should play a vital
role in reduced dimensions \cite{Ando}. Our {\it ab initio} results
show that, indeed, many-electron effects can change qualitatively the
optical spectra of SWCNTs.  Strongly bound exictons are predicted in
small diameter semiconducting nanotubes and even in some metallic
tubes, and they dominate the optical response.

Below, motivated by recent experiments \cite{ZMLi,Bachilo}, we compute
the optical absorption spectra of the three small-diameter SWCNTs:
(3,3), (5,0), and (8,0). We use a recently developed approach in which
electron-hole excitations and optical spectra of real materials are
calculated from first principles in three stages \cite{Rohlfing}: (i)
we treat the electronic ground-state with {\it ab initio}
pseudopotential density-functional theory (DFT) \cite{Sham}, (ii) we
obtain the quasiparticle energies $E_{n{\bf k}}$ within the {\it GW} 
approximation
for the electron self-energy $\Sigma$ \cite{Hybertsen} by solving the
Dyson equation:
\begin{displaymath}
\left[ -\frac{\nabla^2}{2} + V_{ion} + V_{Hartree} + \Sigma(E_{n{\bf k}})\right] 
\psi_{n{\bf k}}= 
E_{n{\bf k}}\psi_{n{\bf k}}~,
\end{displaymath}
and (iii) we calculate
the coupled electron-hole excitation energies $\Omega^S$ and spectrum
by solving the Bethe-Salpeter equation of the two-particle Green's
function \cite{Rohlfing,Strinati}:
\begin{displaymath}
\left(E_{c{\bf k}} - E_{v{\bf k}}\right)A^S_{vc{\bf k}} + 
\sum_{{\bf k'}v'c'} \langle vc{\bf k}|K^{eh}|v'c'{\bf k'}\rangle
A^S_{v'c'{\bf k'}} = 
\Omega^S A^S_{vc{\bf k}}~,
\end{displaymath}
where $A^S_{vc{\bf k}}$ is the exciton amplitude, $K^{eh}$ is the 
electron-hole 
interaction kernel, and $|c{\bf k}\rangle$ and $|v{\bf k}\rangle$
are the quasielectron and quasihole states, respectively.  
We obtain the DFT wavefunctions and eigenvalues by solving the
Kohn-Sham equations within the local density approximation (LDA)
\cite{Sham} using a plane-wave basis with an energy cutoff of 60 Ry.
We use {\it ab initio} Troullier-Martins pseudopotentials
\cite{Martins} in the Kleinmann-Bylander form \cite{KB} ($r_c=1.4$
a.u.).  To compare with experiments in which 4 \AA\ diameter SWCNTs
are grown inside zeolites \cite{ZMLi}, we study the (3,3) and (5,0)
tubes in the experimental geometry with a dielectric background of
$AlPO_4$ \cite{dielectric}. For the (8,0) tube, we work in a supercell
with an intertube separation of at least $9.7$ \AA\ to mimic
experiments on isolated tubes \cite{OConnell,Bachilo}.  In supercells,
due to the long range of the screened Coulomb interaction in
semiconduting tubes, unphysical interactions between periodic images
can lead to deviations from the isolated case.  Hence, we truncate the
Coulomb interaction in a cylindrical geometry for the semiconducting tubes (we
find negligible tube-tube interactions or image effects for metallic
tubes where screening is complete).  Due to depolarization effects in
nanotubes \cite{Ajiki}, strong optical response is observed only for
light polarized along the tube axis ($\hat z$). We only consider this
polarization below.

For the metallic tubes (3,3) and (5,0), we find quasiparticle {\it GW}
corrections to the LDA band energies similar to those in graphite,
namely a $\sim 15\%$ stretching of the LDA eigenvalues away from the
Fermi level ($E_F$) \cite{Fong}. While this result is expected for
large diameter metallic nanotubes which resemble a graphene sheet, we
find it also holds for these small diameter metallic tubes where
curvature effects lead to strong $\sigma-\pi$ hybridization
\cite{Blase}.

Fig. 1a shows the quasiparticle density of states (DOS) for the
metallic (3,3) tube featuring a number of prominent one-dimensional
(1D) van Hove singularities (vHs) near $E_F$. Unlike predictions from
simple tight-binding models \cite{Dress}, these vHs are asymmetric
about $E_F$ due to strong curvature effects. The arrow in the figure
indicates optically allowed low-energy transitions. For the (3,3)
metallic tube, the bands forming the first vHs below $E_F$ and the
second vHs above $E_F$ meet at the Fermi level, but optical
transitions between them are symmetry-forbidden. We calculate
$\epsilon _2(\omega)$ in two ways (see Fig. 1b).  First, we neglect
electron-hole interactions and find the existence of a symmetry gap,
i.e. no electron-hole transition with energy below the prominent peak
at $\hbar \omega = 3.25$ eV. Second, we include electron-hole
interactions and solve the Bethe-Salpeter equation. In general, the
electron-hole interaction kernel has two terms: an attractive direct
term involving the screened Coulomb interaction and a repulsive
exchange term involving the bare Coulomb interaction
\cite{Rohlfing}. Fig. 1b shows that for the (3,3) tube the direct
term dominates: one bound exciton is formed with a binding energy of
$86$ meV and an extent of $\sim 50$ \AA\ along $\hat z$. The
surprising result of having a bound exciton in a metal stems from
having the  symmetry gap,
which is possible for a quasi 1-D system where all {\bf
k}-states have well-defined symmetry. Also, the existence of a {\it
sole} bound exciton is due to the metallic screening (the screening
length is $\approx$ 3.2 \AA): the effective electron-hole interaction
along $\hat z$ resembles an attractive $\delta (z)$ function, and in
1D, the Hamiltonian $H=-\frac{1}{2m^*} \frac{d^2}{dz^2}-|V_0|
\delta(z)$ has a single bound eigenstate.

Fig. 2a shows the quasiparticle band structure for the metallic
(5,0) tube.  According to the band-folding scheme \cite{Dress,Hamada},
this tube should be semiconducting. However, curvature effects
\cite{Blase} lead to strong $\sigma-\pi$ hybridization, forcing a band
to cross $E_F$.  Arrows in the figure indicate optically allowed
interband transitions which give rise to the two peaks, labeled A and
B in the optical spectrum in Fig. 2b. When neglecting electron-hole
interactions, peak B has a lower intensity than A because the
transitions contributing to B do not originate from a band extremum
(vHs) but from the crossing at $E_F$. For this tube, electron-hole
interactions do not bind excitons: while the screening length 
in the (5,0) tube is similar to that of the (3,3)
tube, the symmetry of the bands in the (5,0) tube prohibits direct attraction
between the electron-hole pairs contributing to peaks A and B. Thus
the electron-hole interaction is governed by the repulsive exchange
term. This effect, again, is peculiar to nanotubes: in traditional
semiconductors, the attractive direct term dominates over the
repulsive exchange term. Moreover, when electron-hole interactions 
are included, the
exchange term has a larger effect on peak B and suppresses it greatly.

We now compare our results for the (3,3) and (5,0) tubes to
experiments.  In the  work of Li et al. \cite{ZMLi}, 4 \AA\
diameter SWCNTs were grown inside zeolite channels, 
and three main peaks were found in the measured
absorption spectra (see Table 1).
While 4 \AA\ diameter SWCNTs come in only three chiralities, (3,3),
(5,0) and (4,2), it was not possible to assign directly the specific peaks to
specific tubes experimentally. As shown in Table 1, our results for
the (3,3) and (5,0) tubes are in excellent quantitative agreement with
experiment and provide a concrete identification for two of the
observed peaks.  We conclude that the remaining peak at 2.1 eV is due
to the (4,2) tube (other calculations ignoring electron-hole
interactions point to the same conclusion
\cite{CTChan,Ordejon,42}). Moreover, the many-electron suppression of
peak B in the (5,0) spectrum explains the absence of any observed
feature in the measured spectra at $\approx 2.8$ eV.

We now consider the 6.3 \AA\ diameter semiconducting tube (8,0), in
which we expect even larger excitonic effects.  The (8,0) tube has a
calculated LDA minimum band gap of $0.60$ eV. Quasiparticle
corrections dramatically open the gap to $1.75$ eV. This correction is
significantly larger than those in bulk semiconductors with similar
LDA gaps: we again attribute this to the 1D nature of the SWCNTs which
enhances Coulomb effects (as shown in model calculations \cite{Ando}).

Fig. 3a shows the calculated absorption spectrum for an isolated
(8,0) tube. There are three distinct low-energy peaks (labeled A, B,
and C) in the non-interacting spectrum. When electron-hole
interactions are included, we find far more dramatic excitonic effects
than in the metallic cases: each non-interacting peak gives rise to a
series of visible exciton lines with large binding energies of $0.99$
eV, $0.86$ eV and $1.00$ eV for the lowest-energy excitons ($A'_1$,
$B'_1$ and $C'_1$ respectively). These binding energies are more than
ten times larger than those in bulk semiconductors with similar gaps,
and excitonic effects qualitatively change the spectral
profile. Again, these effects stem from the long range of the screened
Coulomb interaction and the 1D nature of the SWCNTs: e.g., the binding
energy of a 1D hydrogenic system is infinite due to the long-range
Coulomb interaction \cite{Loudon}.  We note that the electron-hole
interaction reverses the relative intensity of the first and second
prominent optical peaks.

Theory predicts that there are two varieties of excitons in the (8,0)
tube: bound excitons with energies below the non-interacting optical
gap (A' and B' series) and resonant excitons with energies above the
non-interacting optical gap (C' series).  Fig. 3b shows the
real-space, electron-hole pair probability distribution $|\Phi({\bf
r}_e,{\bf r}_h)|^2$ as a function of the electron position ${\bf r}_e$
for the photoexcited $A'_1$ bound exciton obtained by fixing the
position of the hole ${\bf r}_h$ (the black star in the figure) on a
carbon $\pi$ orbital. Fig. 3c and 3d
show this correlation more quantitatively for the bound $A'_1$ and the
resonant $C'_1$ exciton: $|\Phi|^2$ is plotted along $\hat{z}$ after
integrating out the electron coordinates in the perpendicular plane.
The extent of the bound part of both excitons is $\sim$ 25 \AA.

Our results for the (8,0) tube are in excellent agreement with the
experimental findings of Bachilo et al. \cite{Bachilo,Weisman}: by
performing spectrofluorimetric measurements on various semiconducting
SWCNTs, with diameters ranging from 0.62 to 1.31 nm and chiral angle
from 3 to 28 degrees, optical transitions were assigned to specific
individual (n,m) tubes.  While their SWCNT samples did not contain the
(8,0) tube, they obtained results for tubes with similar diameter and
chirality and provided fits for the first and second optical
transition energies ($\nu_{11}$ and $\nu_{22}$) as a function of
diameter and chiral angle, which they demonstrate to work well for a
wide range of (n,m) values. Their fits provide a ratio of
$\nu_{22}/\nu_{11}=1.17$ for the (8,0) tube. The traditional
non-interacting $\pi$-orbital only tight-binding model gives a ratio
of 2, and the experimental deviation from 2 has been a puzzle
\cite{Ichida,Liu,Mele}. However, as shown in Table 2, our results for
$\nu_{11}$ and $\nu_{22}$ in the (8,0) tube (peaks $A'_1$ and $B'_1$)
and their ratio are in excellent agreement with the deduced
experimental values.  The deviation of $\nu_{22}/\nu_{11}$ from 2 is a
consequence of both band-structure and many-electron effects: one
needs to include both for a basic and quantitative understanding.

In conclusion, we study the optical absorption spectra of metallic and
semiconducting small diameter SWCNTs and obtain excellent agreement
with available experimental data. We show that electron-hole
interactions (which can be either attractive or repulsive) play a
crucial role, especially for semiconducting tubes, in understanding
experimental results. Large excitonic features for both semiconducting
and metallic tubes are seen to be due to the quasi-1D nature of
SWCNTs, and the manner in which they effect the spectra depends
strongly on the rotational symmetries of the tubes.

This work was supported by the NSF under Grant \#DMR0087088, and the
Office of Energy Research, Office of Basic Energy Sciences, Materials
Sciences Division of the U.S. Department of Energy (DOE) under
Contract \#DE-AC03-76SF00098. Computer time was provided at the DOE
Lawrence Berkeley National Laboratory (LBNL)'s NERSC center. Portions
of this work were performed under the auspices of the DOE by the
University of California Lawrence Livermore National Laboratory (LLNL)
under contract No. W-7405-Eng-48. Collaborations between LLNL and LBNL
were facilitated by the DOE Computational Materials Sciences Network.

\begin{table}
\begin{center}
Table 1: Peak position and optical transitions in 4 \AA\  SWCNTs.
\end{center}
\begin{center}
\begin{tabular}{c| c c} 
Nanotube & Theory & Experiment \cite{ZMLi} \\ \hline
(5,0) & 1.33 eV & 1.37 eV \\
(3,3) & 3.17 eV & 3.1 eV \\
(4,2) & - & 2.1 eV \\
\end{tabular}
\end{center}
\label{Table 1}
\end{table}

\vspace{2cm}

\begin{table}
\begin{center}
Table 2: Lowest two optical transition energies for the (8,0) SWCNT.
\end{center}
\begin{center}
\begin{tabular}{c| c c} 
Transition & Theory & Deduced from experiment \cite{Bachilo,Weisman} \\ \hline
$\nu_{11}$ & 1.55 eV & 1.60 eV \\
$\nu_{22}$ & 1.80 eV & 1.88 eV \\
$\nu_{22}$/$\nu_{11}$ & 1.16 & 1.17  \\
\end{tabular}
\end{center}
\label{Table 2}
\end{table}

\begin{figure}
\resizebox{6.0cm}{!}{\includegraphics{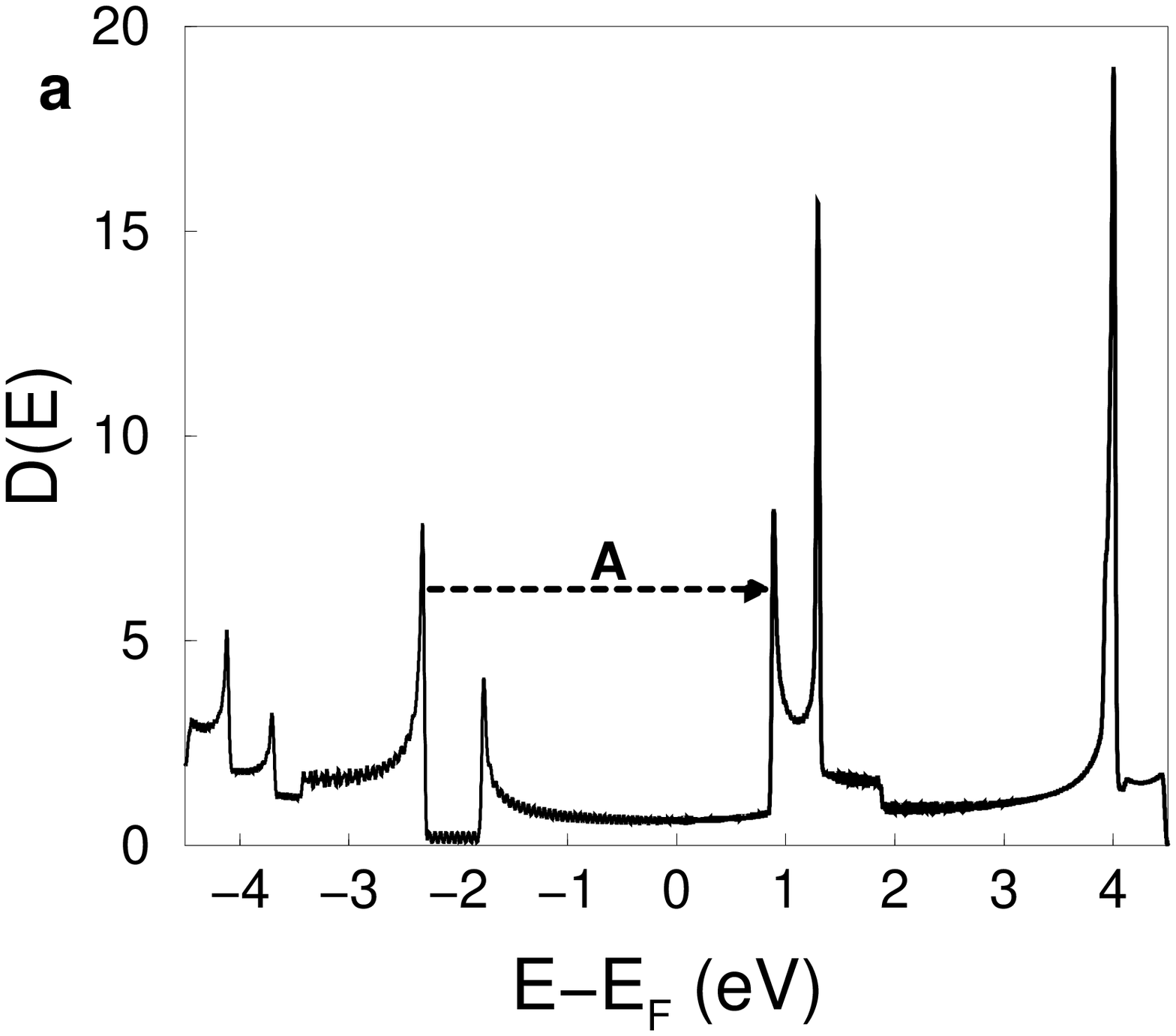}}
\resizebox{6.5cm}{!}{\includegraphics{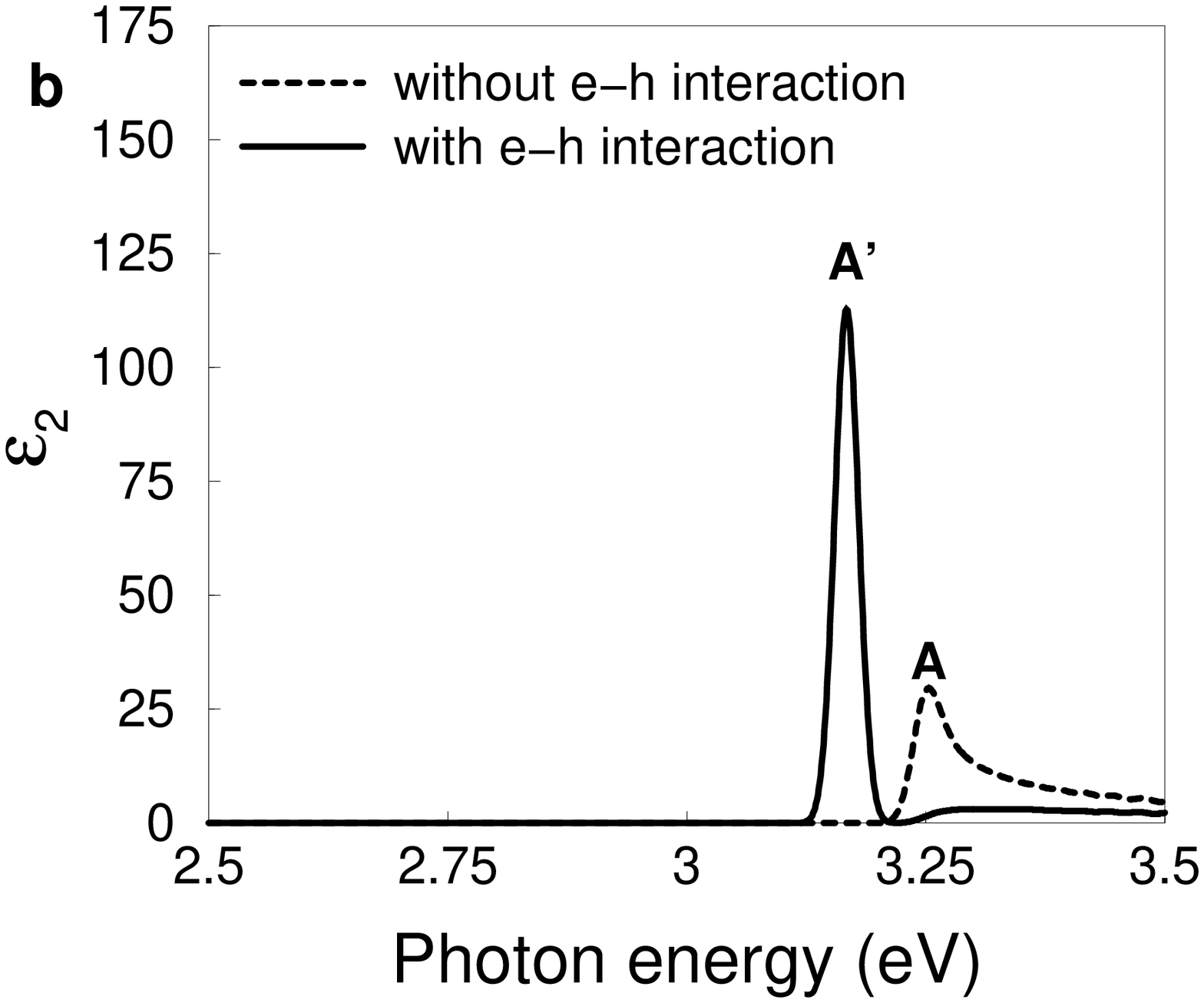}}
\caption{Calculated quasiparticle DOS (a) and absorption spectra (b) 
for the (3,3) SWCNTs in $AlPO_4$ zeolite. Spectra are broadened with a
Gaussian factor of 0.0125 eV.}
\label{Fig1}
\end{figure}

\clearpage

\begin{figure}
\resizebox{6.2cm}{!}{\includegraphics{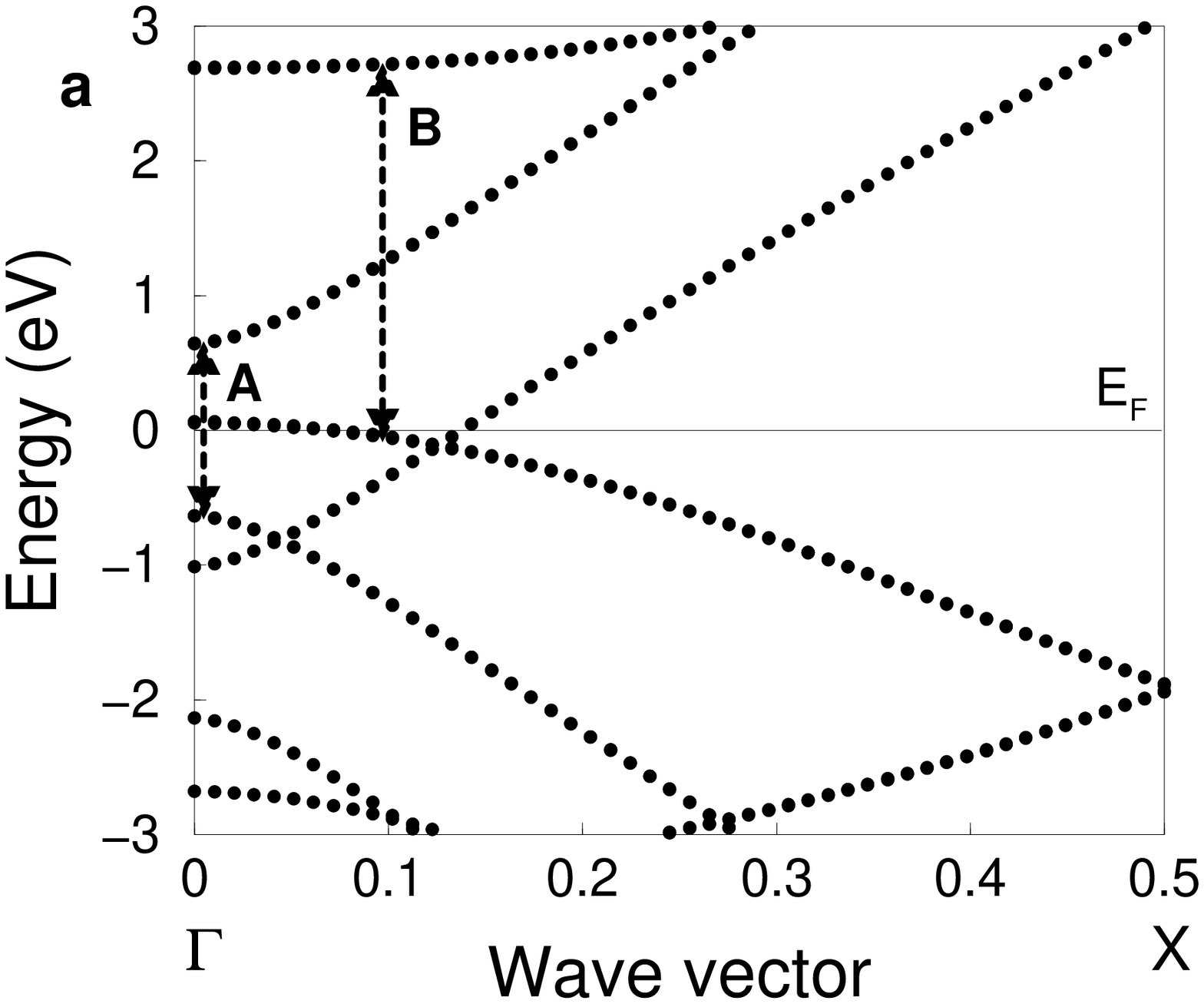}}
\hspace{0.3cm}
\resizebox{6.5cm}{!}{\includegraphics{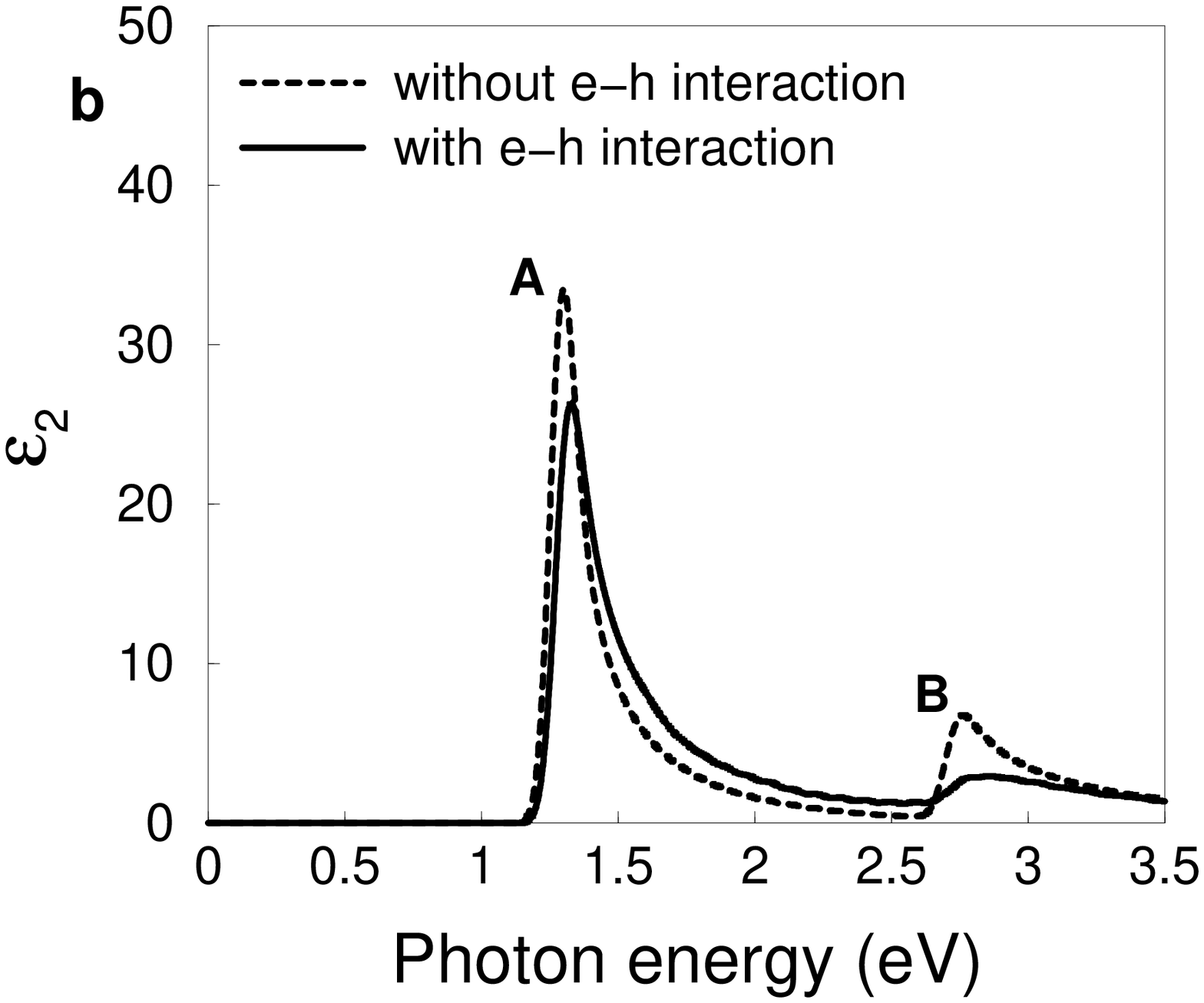}}
\caption{Quasiparticle band structure (a) and absorption spectra (b) for the
(5,0) SWCNTs in $AlPO_4$ zeolite.}
\label{Fig2}
\end{figure}

\clearpage

\begin{figure}
{\resizebox{6.2cm}{!}{\includegraphics{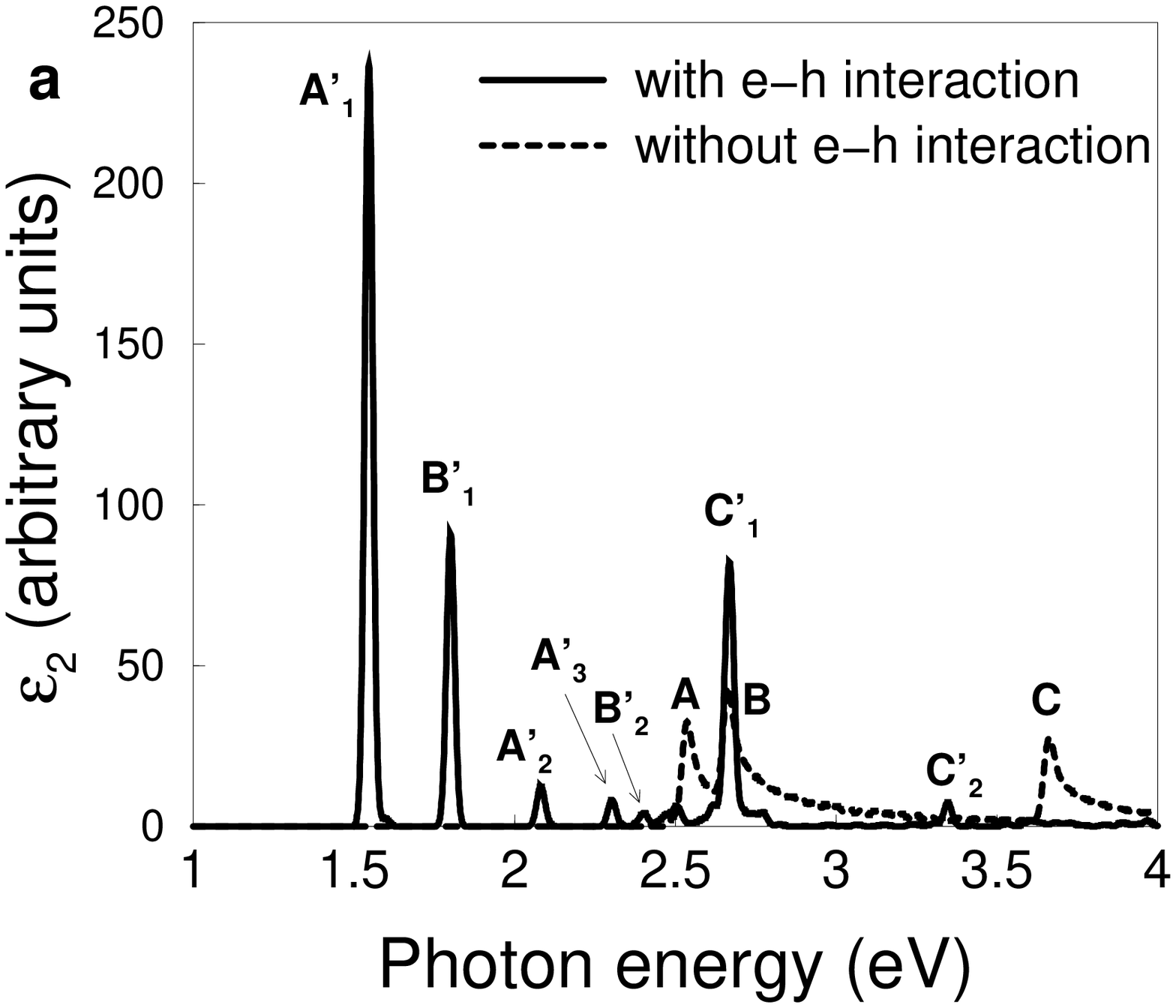}}}
{\resizebox{6.2cm}{!}{\includegraphics{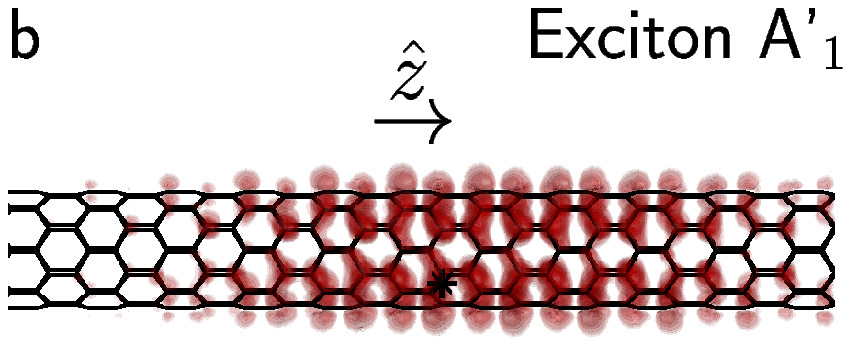}}}
{\resizebox{6.2cm}{!}{\includegraphics{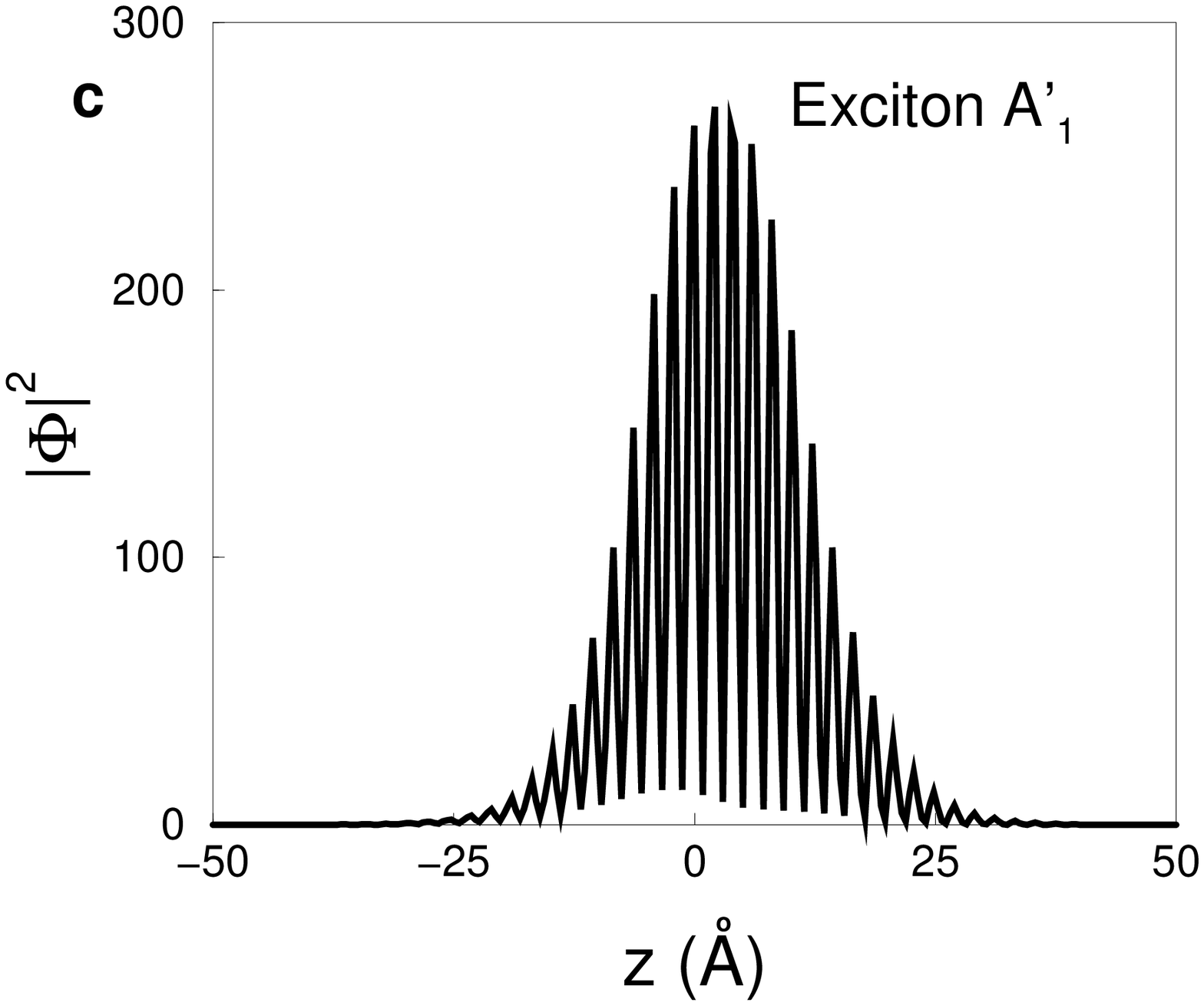}}}
{\resizebox{6.2cm}{!}{\includegraphics{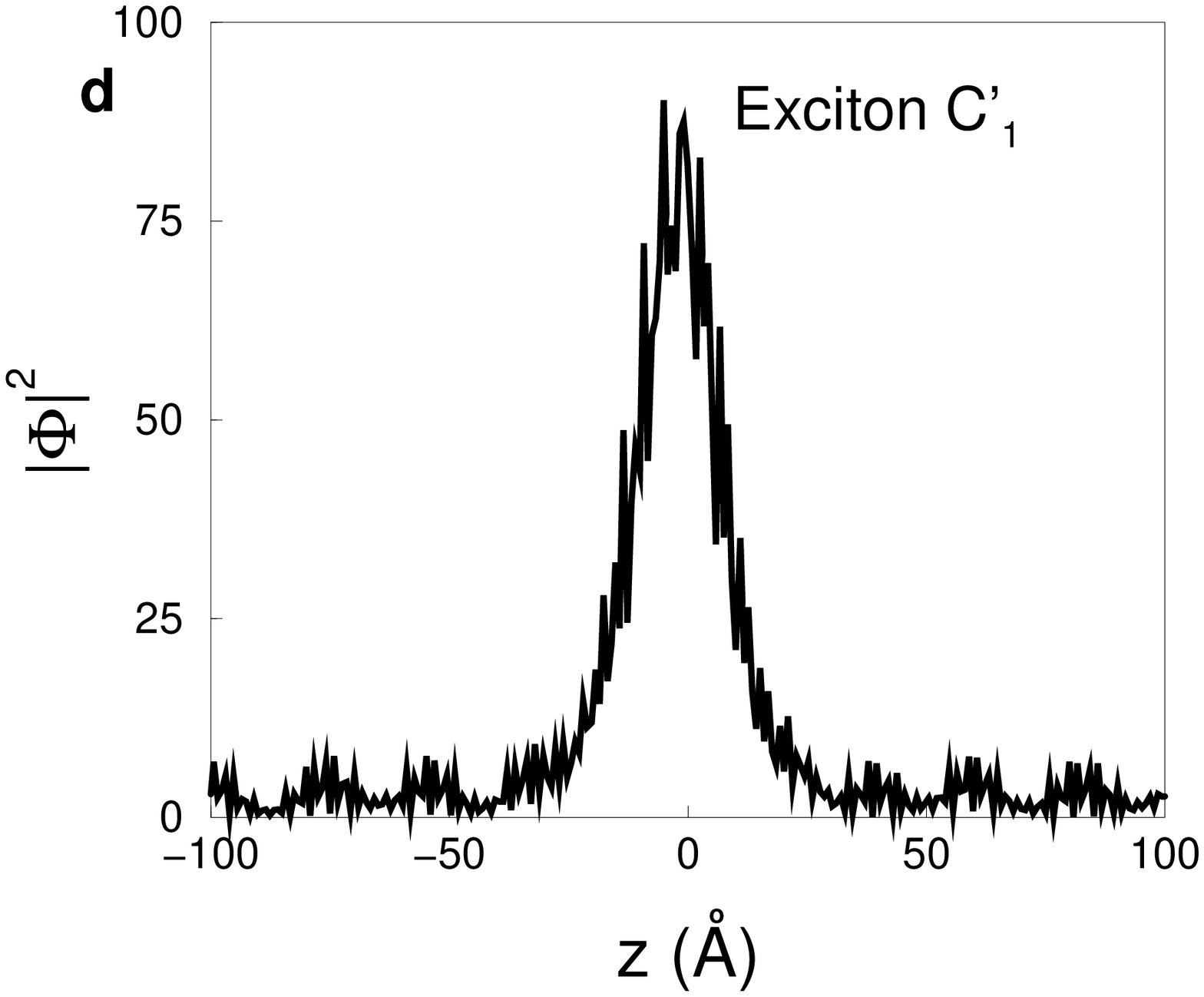}}}
\caption{Absorbtion spectra (a) and exciton wavefunctions (b,c,d) 
for the (8,0) SWCNT. Spectra are broadened with a
Gaussian factor of 0.0125 eV.}
\label{Fig3}
\end{figure}


\begin{references}
\bibitem{dielectric}
The influence of the dielectric background on the optical spectra of
the metallic SWCNTs considered turned out to be negligible.
\bibitem{42}
The calculation for the (8,0) tube (32 atoms in the unit cell)
was already very challenging. We are
currently studying the (4,2) tube (56 atoms in the unit cell),  
but this is a far more difficult calculation. The calculations scale
as the number of atoms to the $4^{th}$ power.
\end{references}
\end{document}